\documentclass[aps,showpacs,twocolumn,prl,floatfix]{revtex4}
\pdfoutput=1

\usepackage{amsmath}
\usepackage{epsfig}
\usepackage{graphicx}
\usepackage{hyperref}
\usepackage{url}

\DeclareMathOperator{\sgn}{sgn}

\newcommand{\beq}{\begin{equation}}
\newcommand{\eeq}{\end{equation}}
\newcommand{\bea}{\begin{eqnarray}}
\newcommand{\eea}{\end{eqnarray}}

\def\leq{\raise 0.4ex\hbox{$<$}\kern -0.8em\lower 0.62ex\hbox{$-$}}
\def\geq{\raise 0.4ex\hbox{$>$}\kern -0.7em\lower 0.62ex\hbox{$-$}}
\def\lsim{\raise 0.4ex\hbox{$<$}\kern -0.8em\lower 0.62ex\hbox{$\sim$}}
\def\gsim{\raise 0.4ex\hbox{$>$}\kern -0.7em\lower 0.62ex\hbox{$\sim$}}
\def\pm{\,\raise 0.4ex\hbox{$+$}\kern -0.8em\lower 0.62ex\hbox{$-$}\,}

\begin{document}

\title{Black Holes are neither Particle Accelerators nor Dark Matter Probes}
\author{Sean T. McWilliams}
\affiliation{Department of Physics, Princeton University, Princeton, NJ 08544}
\email{stmcwill@princeton.edu}
%\author{Sean T. McWilliams \\
%{\small \emph{Department of Physics, Princeton University, Princeton, NJ 08544}}}
%\email{stmcwill@princeton.edu}

\date{\today}

\keywords{black hole physics --- relativity}

\begin{abstract}

It has been suggested that maximally spinning black holes can serve as particle accelerators, reaching arbitrarily high center-of-mass energies.
Despite several objections regarding the practical achievability of such high energies, and demonstrations past and present that such large energies could
never reach a distant observer, interest in this problem has remained substantial.
We show that, unfortunately, a maximally spinning black hole can never serve as a probe of high energy collisions, even in principle
and despite the correctness of the original diverging energy calculation.  Black holes can indeed facilitate dark matter annihilation,
but the most energetic photons can carry little more than the rest energy of the dark matter particles to a distant observer,
and those photons are actually generated relatively far from the black hole where relativistic effects are negligible.  Therefore,
any strong gravitational potential could probe dark matter equally well, and an appeal to black holes for facilitating such
collisions is unnecessary.

\end{abstract}

\pacs{
04.70.Bw, % classical black holes
29.20.-c, % accelerators
95.30.Sf, % relativity and gravitation
95.35.+d, % dark matter
97.60.Lf  % black holes (astrophysics)
}

\maketitle

%\section{Introduction}

\noindent \textbf{\emph{Introduction.}} 
Recently, Ba{\~n}ados, Silk, and West \cite{Ban1} (hereafter BSW) observed an 
interesting feature of the Kerr metric \cite{Kerr} describing the spacetime of
a spinning, uncharged black hole.  Whereas in the case of the Schwarzschild metric describing a nonspinning black hole
\cite{Schwarzschild}, the center-of-momentum (CM) energy $E_{\rm cm}$ for colliding particles asymptotes to a finite limit for
collisions approaching the horizon, BSW showed that $E_{\rm cm}$ diverges for collisions approaching
the horizon of a maximally-spinning (i.~e.~extremal Kerr) black hole.
In a followup paper to BSW \cite{Ban2}, the authors attempted to calculate the flux reaching 
a distant observer from the annihilation of weakly interacting massive particles (WIMPs) in the vicinity of an intermediate
mass black hole (IMBH).  It was argued in BSW that such annihilations could probe Planck scale physics, 
but in \cite{Ban2} the authors made the less ambitious claims that colliding WIMPs around IMBHs could probe ``high energies'', that 
the emission spectrum would likely peak
near the WIMP rest mass, and that a detectable
flux from these collisions could reach reasonable astronomical distances.

Shortly after the appearance of BSW, several practical objections to the proposed mechanism were raised in the literature.  The
authors of \cite{Berti} and \cite{Jacobsen} observed that the BSW effect requires a truly extremal black hole to operate; 
any deviation from extremality reduces $E_{\rm cm}$ to energies of $\mathcal{O}(10\,m_{\chi})$ or less, with $m_{\chi}$ the mass of a
colliding particle.  However, extremal Kerr black holes are thought not to exist in nature, due to the back-reaction from any
infalling radiation \cite{Thorne}.  Similarly, if an extremal black hole somehow managed to exist in perfect vacuum,
a single collision would lower the spin below the extremal limit.  Finally, it was suggested that gravitational radiation, which
had been neglected in the BSW analyses, could significantly decrease the CM energy.  More recently, \cite{Bejger} explicitly
calculated the maximum energy that could reach a distant observer from a collision of infalling particles 
just outside the horizon of an extremal Kerr black hole.
They showed that, while the collision does gain a modest amount of energy through the collisional Penrose process \cite{Penrose},
the gain is only a fraction of the initial rest mass of the colliding particles.

To the objections regarding the physics of the collision in the CM frame, we would add several more
practical limitations.  First, the physics of a Planck-energy collision is completely unknown, but it would
likely result in the formation of a black hole.
If so, the newly formed black hole would either immediately merge with the original black hole, since it obviously will gravitate,
or else it will evaporate, in which case the available energy will be divided among a menagerie of particles.  
Even if beyond-the-standard-model physics somehow avoids black
hole formation, the kinetic energy will still gravitate through 
the nonlinear gravitational interaction of the infalling particles, as conjectured by \cite{Thorne72a}
and demonstrated numerically by \cite{Choptuik}, so the collision product(s) would still likely merge with the original black hole.  
Furthermore, the enormous available energy in these collisions would likely mean that a menagerie of particles would be created and cascade
even without the formation of a black hole.
The branching ratios of such a collision event are completely unknown 
(and possibly unknowable), but the likelihood that such an event would simply yield two photons
seems unlikely.  Also, the cross section for a Planck-energy collision is the Planck scale.  Therefore, the density of particles
necessary to have a possibility of a single collision would be so enormous, that the particles would need to already
be inside a common horizon with the original black hole.  In order to remain outside a common black hole, the density could not be
high enough to facilitate Planck-energy collisions.

Despite all of these practical objections, the BSW problem has drawn substantial interest from the theoretical community.  It was pointed
out in \cite{Jacobsen} and shown explicitly in \cite{Bejger} that the limiting energy reaching an observer is approximately the rest mass of one of the
colliding particles.  However, this result does not negate the proposition that Kerr black holes could be unique probes for dark matter
annihilation, as suggested in \cite{Ban2}, if the peak frequency of the emitted spectrum 
in that case would tell us the mass of the annihilating particles.
Given the possible excess of gamma rays observed in Fermi data at WIMP-scale energies in dark matter subhaloes and
the Galactic center \cite{excess}, the details of annihilation energies around black holes is critical to understand.  Unfortunately, we will show
that, far from probing the Planck scale, collisions occurring near the horizon of an extremal Kerr black hole will generally
emit the lowest energies that reach a distant observer, due to gravitational redshifting.  
The dominant emission will occur at larger distances from the black hole, but still
within the annihilation plateau, and the energy of the emission will indeed be peaked at the WIMP rest mass.  Relativistic effects are therefore
irrelevant for observing dark matter annihilation in the vicinity of black holes, as the black hole serves only as a potential well for
accumulating dark matter at sufficient densities for frequent annihilation.  Therefore, the physics of Kerr black holes, extremal or otherwise, 
plays no role in probing Planck-scale physics, nor does it inform our understanding of the nature of dark matter.

%\section{Kerr geodesics and collisional versus emitted energies}
%\label{sec:main}
\vspace{0.15in}
\noindent \textbf{\emph{Kerr geodesics and collisional versus emitted energies.}}
In \cite{Ban2}, the authors suggested that the collisional Penrose process, whereby a particle extracts energy from the spin of a Kerr black hole
after a collision within the ergosphere, could
enhance the BSW effect.  However, it was recently shown that emitted massive particles can only gain $\sim30$\% of the initial
rest energy of the infalling particles through the collisional Penrose process \cite{Bejger}.
Furthermore, it is known that any electromagnetic emission resulting from the collision can only be enhanced by a small amount through superradiance 
(at most 4.4\% \cite{Teuk}), and constraining the 
collisional geometry can only alter the particle energies modestly and serves to diminish the energy 
of emitted photons \cite{Aguirre}.  Since these effects cannot play a significant role in studying the BSW 
effect, we follow the original approach used in \cite{Ban1,Ban2} and investigate the behavior of ``colliding'' equatorial geodesics,
where the infalling particles follow time-like geodesics until they collide, and their total energy in the CM frame is then imparted
to two null geodesics.

\begin{figure}
\includegraphics[trim = 0mm 0mm 0mm 0mm, clip, width=.45\textwidth, angle=0]{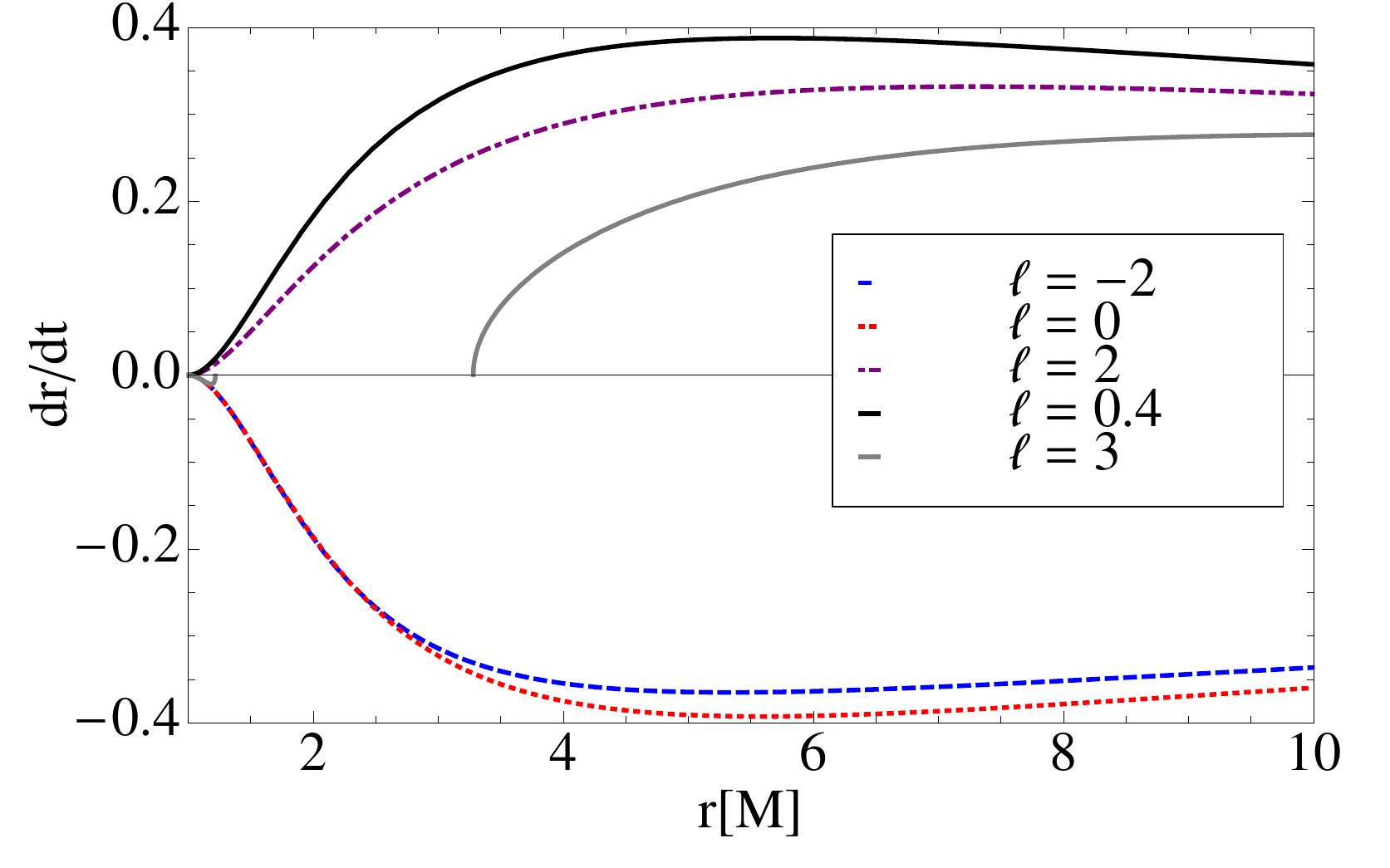}
\caption
{
Radial velocity of several null geodesics as viewed by an observer at infinity.
}
\label{fig:drdt}
\end{figure}

For equatorial geodesics in Kerr spacetime, the equations of motion for the radial position and proper time are given in Boyer Lindquist
coordinates by \cite{Bardeen}
\bea
\frac{dr}{d\tau} &=& \pm \frac{1}{\Sigma} \sqrt{T^2-\Delta (r^2+(l-a)^2)} \label{eq:drdtau} \\
\frac{dt}{d\tau} &=& \frac{1}{\Sigma}\left(-a(a-l)+(r^2+a^2)\frac{T}{\Delta}\right)\,, \label{eq:dtdtau} \nonumber \\
\eea
where $\Sigma \equiv r^2+a^2$, $T\equiv \Sigma-la$, $\Delta\equiv \Sigma-2r$, $l$ is the specific angular momentum of the incident particle,
and $a$ is the dimensionless spin parameter of the black hole; $a=0$ corresponds to a nonspinning Schwarzschild black hole, 
while $a=1$ corresponds to an extremal Kerr black hole.  We note that we are using geometrized units 
in Eqs.~\eqref{eq:drdtau} and \eqref{eq:dtdtau}
and throughout this \emph{Letter}, where $G=c=M=1$.  In Fig.~\ref{fig:drdt},
we show several examples of radial geodesic velocities, as measured by a remote observer.  This figure should compare
to Fig. 3a in BSW, but we note some puzzling differences.  Apart from the meaningfulness of the region $r< 1$ shown in BSW,
the positive value of $\dot{r}$ for radii below, at, and above $r=1$ seems to imply geodesics traveling outward through the horizon.
Also, the nonzero values of $\dot{r}$ at $r=1$ for some geodesics in BSW are difficult to understand, given that the infinite redshifting
at the horizon should cause all derivatives to vanish.  Using Eqs.~\eqref{eq:drdtau} and \eqref{eq:dtdtau}, 
it is straightforward to expand $dr/dt$ around $r=1$, where we find
\beq
\frac{dr}{dt}=\pm \frac{(r-1)^2}{2} + \mathcal{O}\left[\left(r-1\right)^3\right]\,,
\label{eq:drdtrlim}
\eeq
where $\sgn{(dr/dt)}=\sgn{\ell}$.  Therefore, $dr/dt$ should vanish for any $\ell$ as $r\rightarrow 1$.
If we instead assume that $\dot{r}$ is meant to denote $dr/d\tau$ instead of $dr/dt$ in BSW, this
does not resolve the issue, since a corresponding plot of $dr/d\tau$ also fails to agree with Fig. 3a in BSW. 
It is critical that the expressions for the velocity components be correct, since the collision energy in the CM frame
is simply related to the four-velocities of the particles $u^{\mu}$ through \cite{Ban1}
\beq
E_{\rm cm} = \sqrt{2} m_{\chi} \sqrt{1 - g_{\mu\nu} u^{\mu}_{_{(1)}} u^\nu_{_{(2)}}  }\,.
\label{eq:ecomgen}
\eeq 
Because our calculation of geodesics must be carried out correctly in order to calculate
the quantities of primary interest, we show in Fig.~\ref{fig:drdt} that $dr/dt$ behaves as expected for all the geodesics of interest.
Specifically, all cases have vanishing velocities as $r\rightarrow 1$, the $\ell=3$ case is shown to bounce off of the centrifugal barrier,
and geodesics with $-2\,\leq\, \ell\, \leq\, 2$ succeed in falling into or escaping from the near-horizon region, with the critical $\ell=\pm 2$ geodesics
having the minimal $|dr/dt|$ at each radius.  Interestingly, despite the apparent disagreement of the radial velocities
shown in the figures, we ultimately agree with the expression for $E_{\rm cm}$ in BSW.  In terms of the variables we have adopted from \cite{Bardeen},
we find
\begin{widetext}
\beq
E_{\rm cm} = \sqrt{2 + \frac{2}{r^2}\left(\frac{T(\ell_1)T(\ell_2)-\sgn{\ell_1}\sgn{\ell_2}\sqrt{V_r(\ell_1)V_r(\ell_2)}}{\Delta}-(\ell_1-a)(\ell_2-a)\right)}\,, 
\label{eq:ecom}
\eeq
\end{widetext}
where $V_r\equiv T^2-\Delta\left(r^2-(\ell-a)^2\right)$.  Defining the lapse $\alpha \equiv d\tau/dt = \sqrt{-g_{00}}$ and inverting Eq.~\eqref{eq:dtdtau},
we can easily relate the energy observed by a distant observer, $E_{\infty}$, to $E_{\rm cm}$ using 
\beq
E_{\infty}=\alpha E_{\rm cm}\,.
\label{eq:einf}
\eeq
It was observed in \cite{Bejger} that while $E_{\rm cm}$ grows without bound near the horizon, $E_{\infty}$ is bounded to be only marginally
larger than $m_{\chi}$.  However, by expanding Eqs.~\eqref{eq:ecomgen} and \eqref{eq:einf} to leading order in $r-1$,
\bea
E_{\rm cm} &=& 2\sqrt{\frac{4+2\sqrt{2}}{r-1}}+\mathcal{O}\left(\sqrt{r-1}\right)\,, \nonumber \\
E_{\infty} &=& \sqrt{(4+2\sqrt{2})(r-1)^3}+\mathcal{O}\left((r-1)^{5/2}\right)\,,
\label{eq:elim}
\eea
we see that although $E_{\rm cm}$ diverges for $r\rightarrow 1$ for an extremal black hole, $E_{\infty}$ vanishes for small $r-1$.
If we take the formal limit $r\rightarrow 1$, $a=1$, and we maximize the collision energies by choosing $\ell_1=2$ and $\ell_2=-2$
and assuming only one photon escapes, we find $E_{\infty}=m_{\chi}$.  However, the infinite redshift at the horizon means the photon from
this highly tuned collision takes an infinite time to reach the observer.  As soon as we allow $r-1$ to be very small, but nonzero,
we see from Eq.~\eqref{eq:elim} that collisions at smaller radii lead to smaller $E_{\infty}$.  Obviously, the limit $r\rightarrow \infty$
also yields $E_{\infty}=m_{\chi}$, but here the photons actually reach the observer in finite time.  Therefore, the most energetic emission
that a distant observer receives is actually emitted far from the black hole, which is why we have stated that the black hole itself
is a red herring, and any gravitational potential that causes dark matter to collect and annihilate will be just as effective as a dark matter probe.

\begin{figure}
\includegraphics[trim = 0mm 0mm 0mm 0mm, clip, width=.46\textwidth, angle=0]{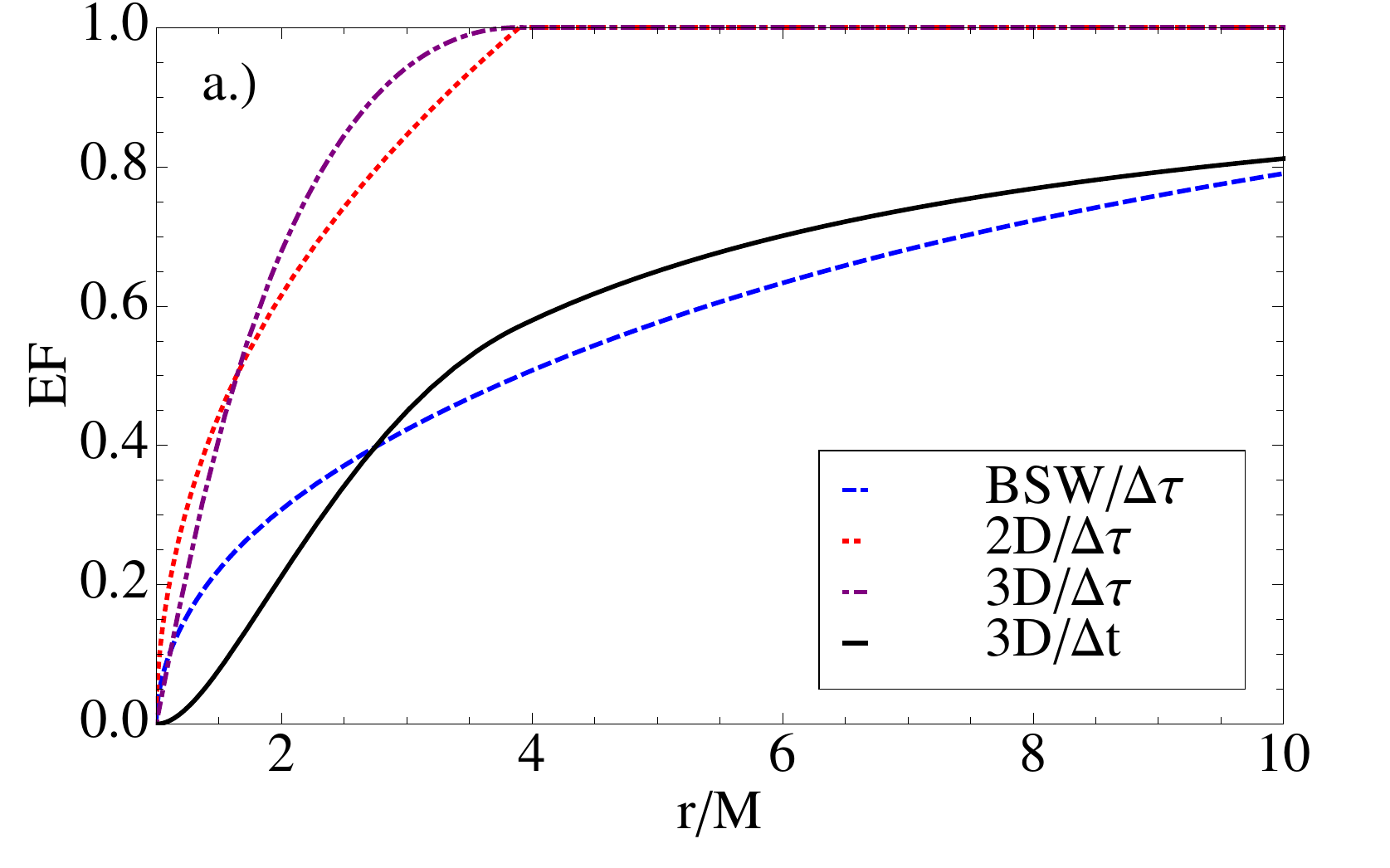}
\includegraphics[trim = 0mm 0mm 0mm 0mm, clip, width=.44\textwidth, angle=0]{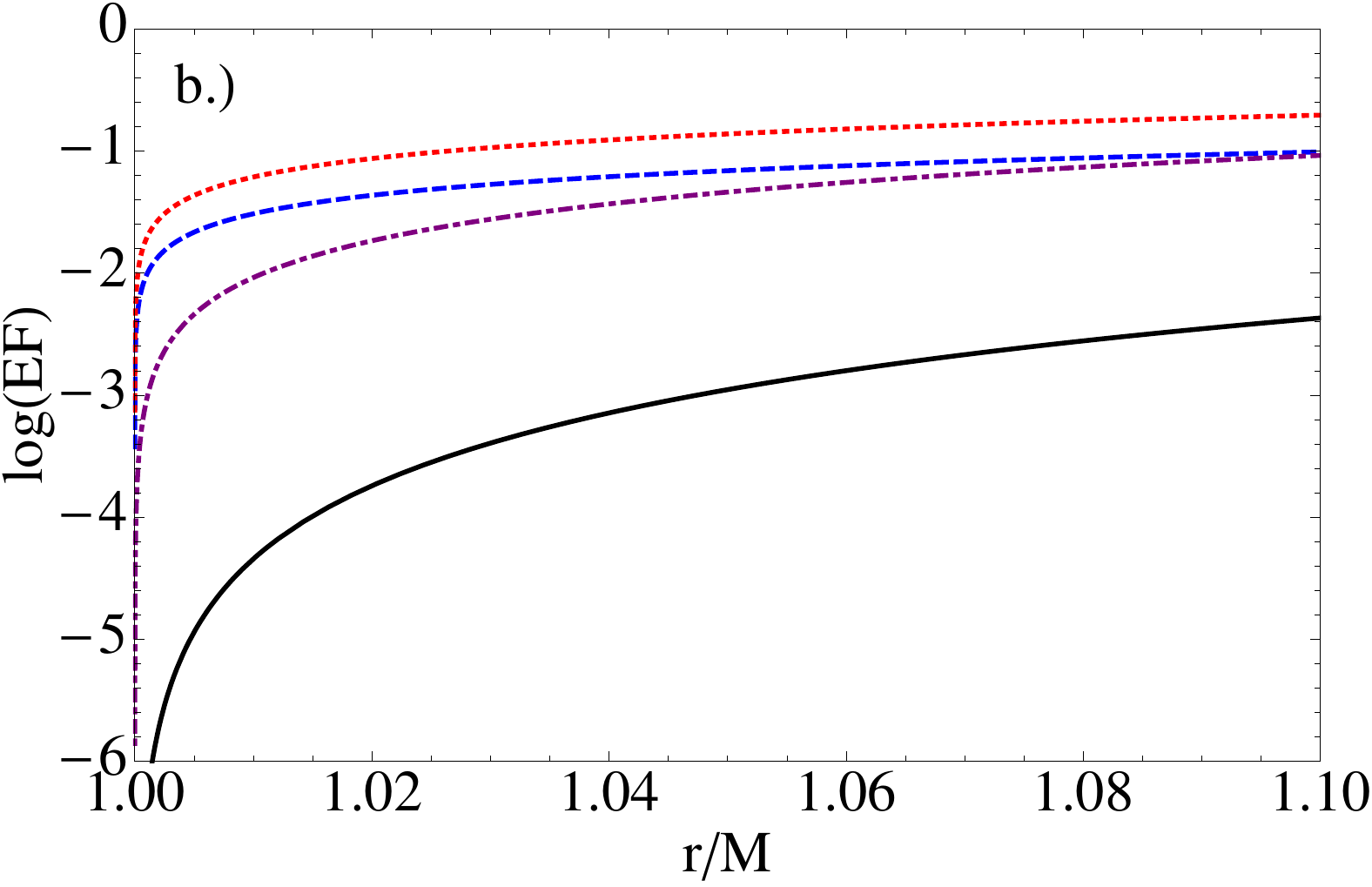}
\caption
{
Escape fraction as a function of radius. a) shows the domain $1\, \leq\, r \,\leq\, 10$, whereas b) shows a zoomed-in view of $1\, \leq\, r\, \leq\, 1.1$.
Where the simple correction of a factor 2 plainly increases the escape fraction relative to \cite{Ban2} by 2 everywhere, the other corrections 
discussed in the text greatly diminish the escape fraction for $r \, \leq \, 1.1$, so 
that our calculated escape fraction is several orders of magnitude below that calculated
in \cite{Ban2} for small radii.
}
\label{fig:EF}
\end{figure}

Equipped with the quantities necessary to calculate the properties for any equatorial collision, we first proceed 
to calculate the escape fraction for products of the collision as a function of the distance between the collision and the center of the black hole
in Boyer-Lindquist coordinates.  Specifically, we follow the procedure given in detail in Sec. III of \cite{Ban2} for calculating the range of
angles around the perpendicular, $\Delta\phi$, inside of which an emitted photon would escape to infinity (see also \cite{Bejger} and Box 25.7 of \cite{MTW}).  In using
this result to calculate the overall fraction of collision-produced photons that can escape, we make two critical corrections to the calculation in \cite{Ban2}.
In \cite{Ban2}, the authors justifiably assume that infalling particles are confined to the equatorial plane of the black hole.  They further
assume that the emitted particles are also constrained to the equatorial plane (as do the authors of
\cite{Bejger}, incidentally).  However, this latter assumption is unjustified; since
two photons will be emitted in the optimal case, their momenta out of the equatorial plane is constrained to be equal and
opposite, but need not be zero.  We define $\Delta\phi$ as in \cite{Ban2}, to be the angle in the CM frame
between the unit normal to the horizon within the equatorial
plane at the point of collision, and the trajectory corresponding to the maximum (minimum) specific angular momentum for a null geodesic to escape,
$\ell = 2(-2)$.  Assuming a collision with no net angular momentum, the authors of \cite{Ban2} assume the emission is equally likely
to occur across the full $2\pi$ of the equatorial plane in the CM frame, so that the escape fraction $EF$ is given by
%so that the horizon subtends all of the equatorial plane except the sliver $\Delta\phi$.
\beq
EF_{\rm BSW}=\Delta\phi/2\pi\,.
\label{eq:efbsw}
\eeq
Given the definition of $\Delta\phi$, emission artificially constrained to the equatorial plane should actually have $EF_{\rm 2D}=2EF_{\rm BSW}$.
However, more critically, the correct escape fraction is given by the ratio of the area of the sky circumscribed by a cone with apex angle $\Delta\phi$
to the full $4\pi$ area,
\beq
EF_{\rm 3D}=\frac{1}{2}\left(1-\cos \Delta\phi\right) = \left(\frac{\Delta\phi}{2}\right)^2 + \mathcal{O}\left(\Delta\phi^4\right)\,.
\label{eq:ef3d}
\eeq

Finally, if we wish to calculate the observed flux from the number of available infalling particles as done in \cite{Ban2}, we must calculate the escape fraction 
over some time interval.  Here again we make a correction to the approach in \cite{Ban2}, which implicitly assumes that the rate of emitted particles
observed in the CM frame is the same as the rate observed by a distant observer.  However, time runs far more slowly for the CM observer sitting just outside 
of the black hole horizon.
Since the quantity of interest is the flux reaching a distant observer, any rate calculation
should use the time measured by the observer, rather than the time observed in the CM frame of the collision.  The flux 
calculated in \cite{Ban2} must therefore be corrected by the lapse $\alpha$, given by the inverse of Eq.~\eqref{eq:dtdtau}, 
which relates the rate of clocks in the CM frame to the rate of clocks at an infinite distance, so that the observed flux is given by
\beq
\Phi = \alpha \Phi_{\rm BSW}\,,
\label{eq:flux}
\eeq
where, for convenience, we note that 
\bea
\alpha(r,a,\ell) &=& \frac{\left(r^2+a^2\right)\left(r^2-2r+a^2\right)}{r\left[r^3+a^2r+2a(a-\ell)\right]} \nonumber \\
\alpha(r,1,\ell) &=& \frac{(r-1)^2(r^2+1)}{r(r^3+r-2\ell+2)}\,.
\label{eq:dtaudt}
\eea

In Fig.~\ref{fig:EF}, we compare the various estimates of the escape fraction divided by a 
fixed reference time measured either in the CM frame ($\Delta \tau$) or by a distant observer ($\Delta t$), since this combined quantity is closely related
to the flux of photons that an observer will measure.  We note that both $EF_{\rm 2D}$ and $EF_{\rm 3D}$ reach unity for photons emitted
from $r\, \geq \, 4$, which is expected for photons emitted beyond the centrifugal barrier of the Kerr potential, and which
is contrary to the escape fraction calculated in \cite{Ban2}.  The first correction relative to \cite{Ban2} is a trivial factor of 2 overall,
but the correction from a 2D escape pie slice to a 3D escape cone and the correction to account for the slowing of clocks near black holes both serve to greatly
diminish the escape fraction at small radii by several orders of magnitude.  Therefore, we see that not only do the largest $E_{\rm cm}$ collisions have
the smallest $E_{\infty}$, but they are also the least likely collisions to emit a photon that will eventually reach an observer.

\begin{figure}
\includegraphics[trim = 0mm 0mm 0mm 0mm, clip, width=.45\textwidth, angle=0]{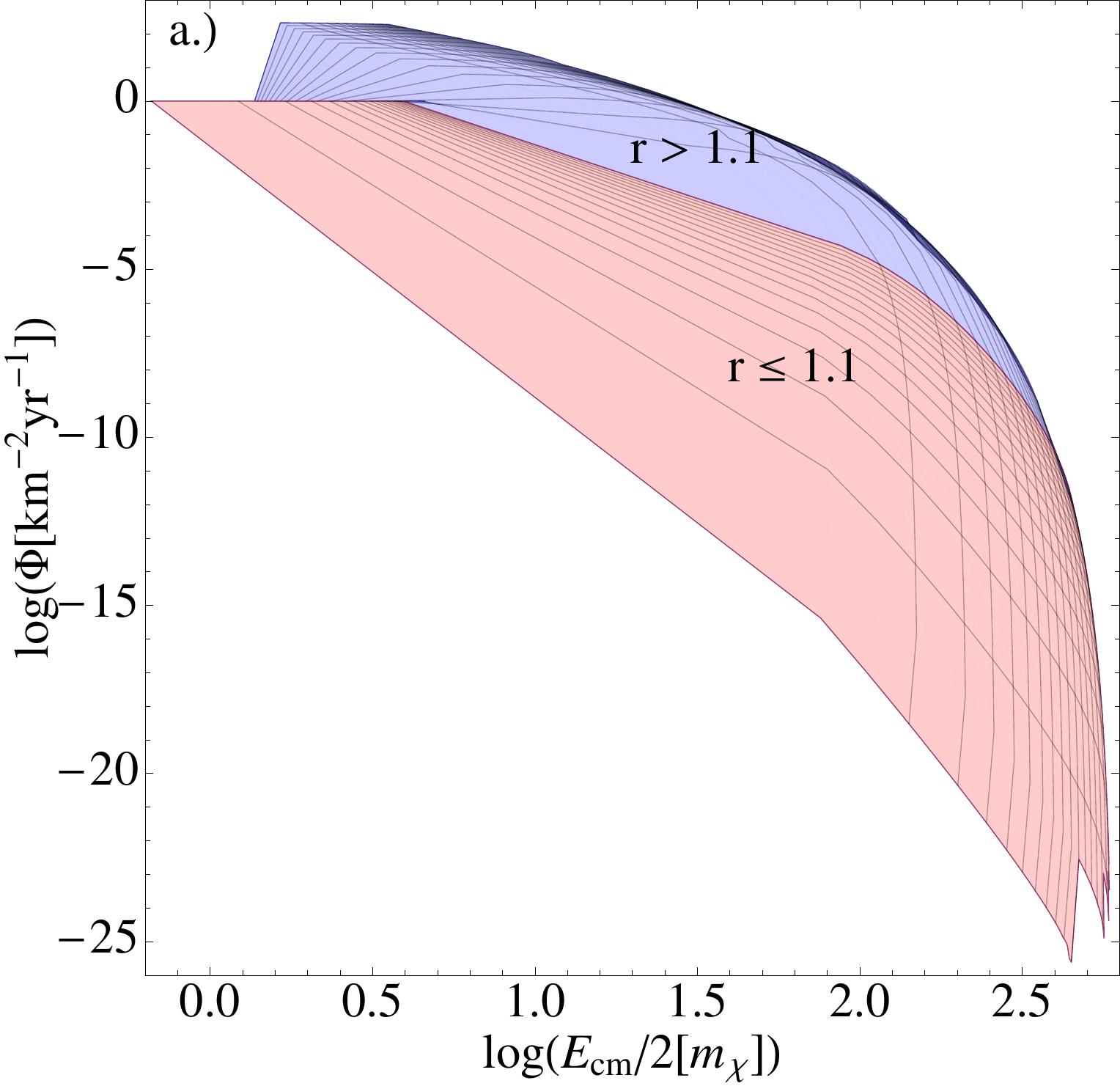}
\includegraphics[trim = 0mm 0mm 0mm 0mm, clip, width=.45\textwidth, angle=0]{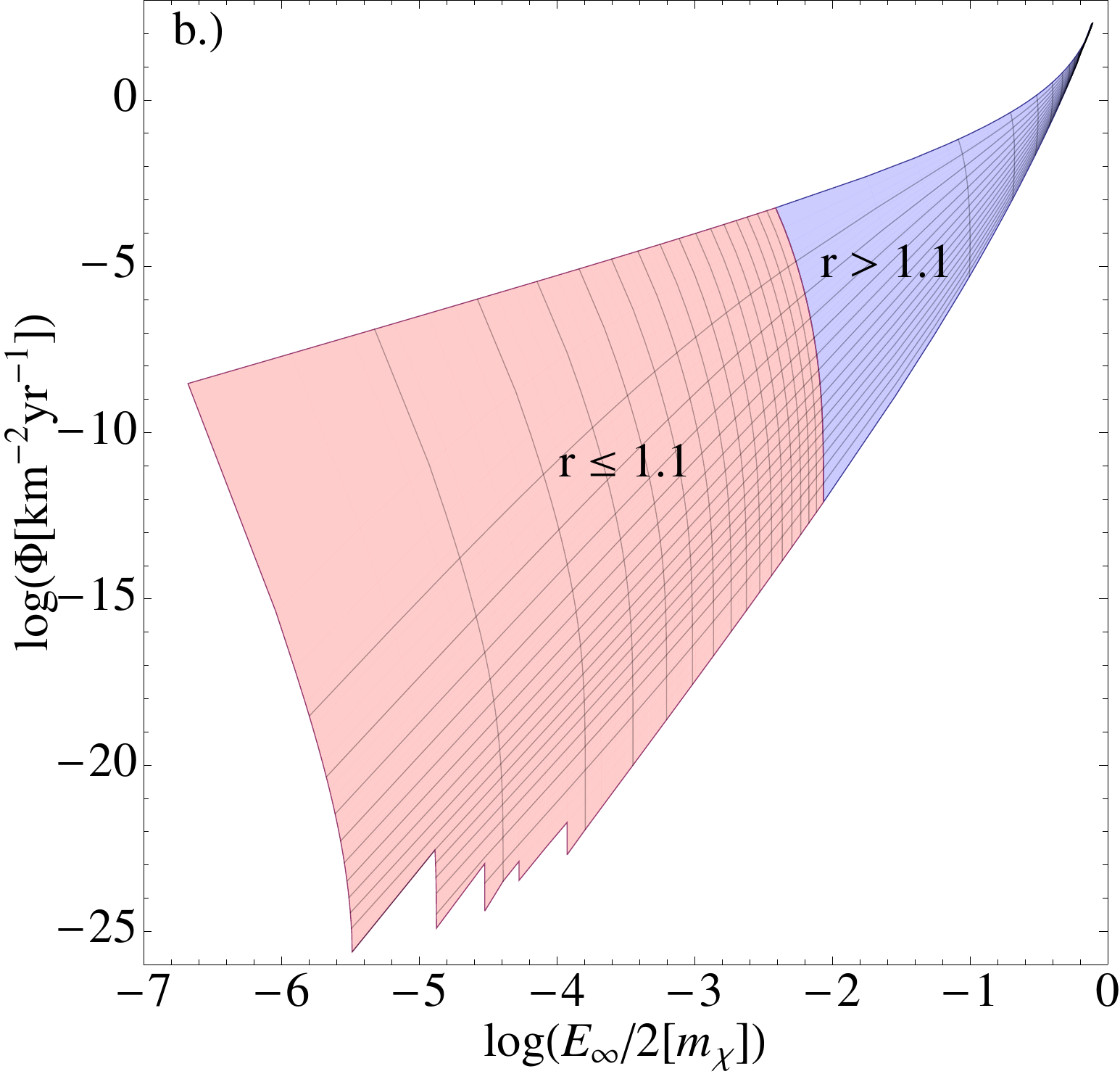}
\caption
{
Contours of energy and flux for $0\, \leq\, \ell\, \leq\, 2$ and $1 \,\leq\, r \,\leq\, 10$.  a.) shows the
CM energy and
b.) shows the energy measured at infinity versus the flux reaching an observer
10 kpc away, assuming the outgoing photon carries half of the collision energy.
}
\label{fig:flux}
\end{figure}

Given how diminished the true escape fraction is compared to \cite{Ban2}, it is instructive to calculate the observed flux of photons and the 
collisional or observed energies for the full range of allowable infalling 
momenta and for collision locations spanning from just outside the black hole horizon at $r=1+\epsilon$ ($\epsilon \ll 1$) out to $r=10$,
which we show in Fig.~\ref{fig:flux}.  
We again note that the authors of \cite{Ban2} calculate the flux emitted from the near-horizon
region {\it as viewed from the same location}.  However, the quantity of interest is the flux reaching a distant observer, so
we instead calculate the flux $\Phi$ that reaches an observer at a distance of 10 kpc.  It is clear from Fig.~\ref{fig:flux}
that the collisions with the largest $E_{\rm cm}$, which will lie in the contours with $r \,\leq\, 1.1$, have a vanishingly small flux
above $E_{\rm cm} \approx 5 m_{\chi}$, so energies much higher than the particle's rest mass are never probed.  The particles
with $r \,\leq\, 1.1$ that do manage to reach an observer actually have $E_{\infty} < m_{\chi}/100$.  As anticipated, we see that all of the photons
with $E_{\infty} \approx m_{\chi}$ that reach a distant observer come from $r>1.1$, so the large $E_{\rm cm}$ just outside the horizon
has no observational relevance.

%\section{Conclusions}
\vspace{0.15in}
\noindent \textbf{\emph{Conclusions.}}
We have shown that, while particle collisions just outside the horizon of an extremal Kerr black hole can indeed reach arbitrarily high
energies, the products of those collisions never reach an observer.  Employing the extremely optimistic assumption that the collisions
always produce only two photons, we have shown that the photons created near the horizon that do reach a distant observer have relatively
low energies $E_{\rm cm} \sim \mathcal{O}(m_{\chi})$ and $E_{\infty} < m_{\chi}/100$, and the most energetic photons to reach
an observer have $E_{\infty} \sim \mathcal{O}(m_{\chi})$ and originate far from the black hole.  Therefore, extremal Kerr black holes
do not provide a means to observe Planck scale physics, nor do they provide a unique probe for annihilating dark matter.

\end{document}